\documentclass[pre,twocolumn,superscriptaddress,floatfix]{revtex4}
\usepackage{graphicx}
\begin{document}
\title[Fast heat transfer calculations...]{Fast heat transfer calculations in supercritical fluids versus hydrodynamic
approach}
\author{V. S. Nikolayev}\email[email:]{vnikolayev@cea.fr}
\affiliation{DSM-DRFMC-Service des Basses Temp\'eratures/ESEME, CEA-Grenoble, France} \altaffiliation[Mailing
address: ] {CEA-ESEME, Institut de Chimie de la Mati\`ere Condens\'{e}e de Bordeaux, 87, Avenue du Dr. A.
Schweitzer, 33608 Pessac Cedex, France}
\author{A. Dejoan}
\altaffiliation[Present address: ]{Aeronautics Department, Imperial College, Prince Consort Road, South
Kensington, London, England, SW7 2BY,  United Kingdom}
\author{Y. Garrabos}
\affiliation{CNRS-ESEME, Institut de Chimie de la Mati\`{e}re Condens\'{e}e de Bordeaux,\\
87, Avenue du Dr. A. Schweitzer, 33608 Pessac Cedex, France}
\author{D. Beysens}
\affiliation{DSM-DRFMC-Service des Basses Temp\'eratures/ESEME, CEA-Grenoble, France}
\date{\today}

\begin{abstract}
This study investigates the heat transfer in a simple pure fluid whose temperature is slightly above its
critical temperature. We propose a efficient numerical method to predict the heat transfer in such fluids
when the gravity can be neglected. The method, based on a simplified thermodynamic approach, is compared with
direct numerical simulations of the Navier-Stokes and energy equations performed for CO$_2$ and SF$_6$. A
realistic equation of state is used to describe both fluids. The proposed method agrees with the full
hydrodynamic solution and provides a huge gain in computation time.  The connection between the purely
thermodynamic and hydrodynamic descriptions is also discussed.
\end{abstract}\maketitle

\section{Introduction}

In fluids near their liquid-gas critical point, the characteristic size of the density fluctuations becomes
larger than the characteristic size of the molecular structure.  Consequently, the fluid behavior is ruled by
fluctuations and not by its particular molecular structure. This implies that most fluids behave similarly
near the critical point (like the 3D Ising model). This universality makes the study of near critical fluids
very appealing.  Due to their very low thermal diffusivity and to their very large thermal expansion and
compressibility, the study of heat transfer in such fluids is particularly challenging.  When such a fluid is
confined in a heated cavity, a very thin hot boundary layer develops and induces a fast expansion that
compresses the rest of the fluid.  The resulting pressure waves spread at the sound velocity (i.e very
rapidly) and adiabatically compress the bulk of the fluid which is therefore homogeneously heated. After
several sound wave periods the pressure is already equilibrated and can be assumed to be nearly homogeneous
along the cavity.  During the initial stage of heating, this process of energy transfer, called ``Piston
effect" \cite{PE-Zap,Onuki}, is much more efficient than the usual diffusion scenario. Indeed, if the Piston
effect were absent, the bulk of the fluid would remain at the initial temperature.

In the industrial domain, the Piston effect can be used to transfer heat much faster than by conduction. This
feature can be readily applied to the development of heat exchangers under microgravity conditions
\cite{Prop,Caloduc} where heat transfer by natural convection is obviously not possible.

While the physical origin of the Piston effect is well understood, the calculations required to represent
realistic experimental conditions are difficult because the inherent nonlinear dynamical behavior of such
fluids is complicated by the highly non-linear equations of state (EOS) used to describe real near-critical
fluids.  Two computational approaches have been suggested for confined fluids in absence of convection. In
one of them, which we will refer as the ``thermodynamic approach" \cite{Bouk}, the Piston effect is taken
into account by a supplementary term $g$, introduced in the heat conduction equation as follows,
\begin{equation}\label{hcond}
    {\partial T\over\partial t}={1\over\rho c_p}\nabla\cdot(k\nabla T)+g(T),
\end{equation}
with
\begin{equation}\label{g}
    g(T)=\left(1-{c_v\over c_p}\right)\left({\partial T\over\partial p}\right)_\rho{\partial p\over\partial t},
\end{equation}
where $T$ is the local fluid temperature, $c_v(c_p)$, the specific heat at constant volume (pressure) per
unit mass, $\rho$ the local density and $k$ the thermal conductivity of the fluid.  One can note that the
term $g(T)$ is only relevant near the critical point where $c_p\gg c_v$.

The fluid motions are neglected and the pressure $p$,
assumed to be homogeneous, is only a function of time $t$.
The pressure $p$ is determined from the fluid mass conservation and
computed via the nonlinear expression \cite{Bouk}
\begin{equation}\label{pB}
    {\partial p\over\partial t}=-{\int_v(\partial\rho/\partial T)_p\;\partial T/\partial t\, \mathrm{d}v\over
    \int_v\rho\chi_T\;\mathrm{d}v}
\end{equation}
where $\chi_T=\rho^{-1}(\partial\rho/\partial p)_T$ is the isothermal
compressibility and $v$ the volume of the fluid sample.  The
resolution of (\ref{pB}) requires an iterative procedure for each time
step. This consists in calculating the temperature \emph{in the whole
fluid volume} using Eqs.~(\ref{hcond}-\ref{g}) for some trial value of
$p$ (and thus $\partial p/\partial t$) and determining the other
thermodynamic parameters ($\rho$, $\chi_T, \ldots$) by an EOS
\begin{equation}\label{EOS}
\Lambda(p,\rho,T)=0.
\end{equation}
Subsequent computation of the \emph{volume} integrals in Eq.(\ref{pB}) gives a new value of $\partial
p/\partial t$ from which the pressure $p$ is corrected. The correction step is repeated until convergence.
This approach has been extensively used by several groups \cite{Straub,Zhong} in one-dimensional (1D)
calculation in conjunction with the restricted cubic EOS \cite{Seng} and the finite difference numerical
method.  However, its extension to higher dimensions induces a large computational effort.  First, it
requires a sophisticated programming and, second, the computer resources rise steeply because the
thermodynamic variables have to be evaluated at each grid point of the computational domain by means of the
iterative procedure to solve Eq.~(\ref{pB}).

Teams with expertise in theoretical hydrodynamics have developed a rigorous hydrodynamic approach
\cite{PE-Zap} in which the Navier-Stokes and energy equations are coupled with the EOS. This set of equations
has been solved analytically in 1D by asymptotic matching techniques \cite{HZAPCAR} and also by direct
numerical simulation (DNS) \cite{HAMIOUA} via the finite volume method \cite{HPAT} both in 1D and 2D. For the
sake of simplicity, the van der Waals EOS has been used in all these works. Although this simple EOS provides
satisfactory qualitative results, accuracy can only be ensured by using a realistic EOS. However, as shown in
the present study, inserting a realistic EOS into the hydrodynamic equations leads to a much more difficult
computational task, which involves prohibitively large calculation times.  For instance, to reach the final
steady state in one of the 1D runs carried out for the cubic EOS in the present study, one requires about two
months on a 800 MHz PC.

The purpose of this article is twofold. We first formulate an
approximate method which is both simple and rapid (e.g., the same
calculation cited above required only 20s !)  and then compare it
with the DNS formalism.  Second, since this work is the first one
to use a realistic EOS for the DNS, we describe in detail the
hydrodynamic approach for the near-critical fluids with a general
EOS. We expect that this complete and unified description may be
useful for the scientific community, as the hydrodynamic method is
dispersed over many (some of them not easily accessible) publications.

The article is organized as follows. In Sec. \ref{meth} we describe
the fast calculation method.  Sec.~\ref{Anne} presents the
hydrodynamic approach. Sec.~\ref{Results} deals with the comparison
between both approaches. The conclusions are given in
Sec.~\ref{Concl}.

\section{ Fast Calculation Method}\label{meth}

The method is based on the thermodynamic approach, i.e. on the energy equation~(\ref{hcond}). However, a
different pressure equation will be used instead of Eq.~(\ref{pB}). While the latter equation integrates over
the fluid \emph{volume}, we are looking for a pressure equation that integrates only over the
\emph{boundaries} of the fluid domain. Such an equation would accelerate the iterative procedure.  However,
its advantage would not be decisive without an appropriate numerical method for Eq.~(\ref{hcond}) that should
only require computation of the thermodynamic variables at the boundaries. Such a numerical method is called
Boundary Element Method (BEM) and is broadly used in heat transfer problems. The BEM is expounded in
Appendix~\ref{appA}.

Several simplifications have to be introduced before
presenting the formulation of the
energy and pressure equations used in the
present method. As stated in the introduction,
we are mainly interested
in the description of the early stages of the heating,
i.e. on times of the order of the Piston effect time scale
\cite{HGARBON}. In this regime, the thermodynamic quantities
$c_p,c_v,k,\ldots$ are constant in the bulk of the fluid and
only vary in the very thin hot and cold
boundary layers. Thus, the following assumptions can be made:

(i) The spatially varying parameters ($c_p,c_v,k$, etc.) in Eq.~(\ref{hcond}) can be replaced by the
spatially homogeneous time dependent values which will be denoted hereafter by an over-bar (e.g.
$\bar{c}_p$). These values are calculated with the EOS using density $\langle\rho\rangle$ (where the brackets
indicate the spatial average)  and the pressure $p=p(t)$. We note that the density $\langle\rho\rangle$ is
constant as the system is closed.

This assumption is equivalent to the statement that all these
quantities should be calculated using the temperature value $\bar
T=\bar T(t)$ obtained from the EOS for the pressure $p$ and
the density $\langle\rho\rangle$. The quantities calculated in such a way
correspond to the over-bar variables mentioned above . In general,
for a thermodynamic quantity $X$,
$\bar X\ne\langle X\rangle$.

(ii) During the system evolution, the thermodynamic quantities are supposed
not to vary sharply in time.

(iii) The initial temperature $T_0$ is uniform.

The assumption (iii) is made for simplicity and can be relaxed if necessary. However, the assumptions (i) and
(ii) are essential for this approach.

In the following section we formulate the energy and pressure
equations that govern the kinetics of the supercritical fluid in a
reduced gravity environment.

\subsection{Energy equation}

Using the assumption (i) and the constancy of $\langle\rho\rangle$, one can write
\begin{equation}\label{dp}
\textrm{d}p=\overline{(\partial p/\partial T)}_\rho\textrm{d}\bar T
\end{equation}
so that the term (\ref{g}) reduces to
\begin{equation}\label{gb}
    \bar g(\bar T)=\left(1-{\bar{c}_v\over \bar{c}_p}\right){\textrm{d} \bar T\over \textrm{d} t},
\end{equation}
and Eq. (\ref{hcond}) can be reduced to the equation
\begin{equation}\label{hb}
    {\partial \psi\over\partial t}=\bar D\nabla^2 \psi.
\end{equation}
The thermal diffusion coefficient $\bar D=\bar{k}/\langle\rho\rangle \bar{c}_p$ depends on $\bar T$ i.e. on
time $t$ only, and
\begin{equation}\label{psi}
  \psi(\vec{x},t)=T(\vec{x},t)-T_0-E(\bar T).
\end{equation}
with $\vec{x}$ the  position vector and
\begin{equation}\label{E}
    E(\bar T)=\int\limits_{T_0}^{\bar T} \left(1-{\bar{c}_v\over \bar{c}_p}\right)\textrm{d}\bar T.
\end{equation}
The initial condition is  $\psi|_{t=0}=0$ because according to the assumption (iii),
\begin{equation}\label{initT}
 \bar T|_{t=0}=T_0.
\end{equation}
Finally, a known dependence of $\bar D=D_d\,f(\bar T)$, where $D_d$ is a dimensional constant and $f$ is a
non-dimensional function, allows the time $t$ to be replaced by a new independent variable $\tau$ defined by
the equation
\begin{equation}\label{tau}
 {\textrm{d}\tau\over\textrm{d}t}=f(\bar T),  \\
\end{equation}
whose initial condition can be imposed as $\tau |_{t=0}=0$. Since $\bar T$ is a function of $t$ only, this
initial value problem is fully defined. The substitution of Eq.~(\ref{tau}) into Eq.~(\ref{hb}) results in
the linear diffusion problem with the constant diffusion coefficient $D_d$
\begin{equation}\label{hb1}
\begin{array}{r}\displaystyle
  {\partial \psi\over\partial \tau}=D_d\nabla^2 \psi,\\
  \psi|_{\tau=0}=0.
\end{array}
\end{equation}
It can be solved with BEM as shown in Appendix~\ref{appA}.

Usually, the "temperature step" boundary condition has been applied for 1D problems. This heating process
corresponds to a fluid cell, initially at a uniform temperature, which is submitted to a sudden increase of
temperature at one of its boundaries, while the other is kept at the initial temperature (Dirichlet boundary
conditions). This heating condition is physically unrealistic because the initial value for the heat flux at
the heated boundary is infinite. Instead, in this work we use Neumann-Dirichlet boundary conditions: a heat
flux $q_{in}$ is imposed at one of the boundaries, while the initial temperature $T_0$ is maintained at the
other boundary.

\subsection{Boundary form of the pressure equation}

Let us begin by writing the linearized relationship, valid under the assumption (ii):
\begin{equation}\label{dr}
    \delta\rho=\left({\partial\rho\over\partial s}\right)_p\delta s+
    \left({\partial\rho\over\partial p}\right)_s\delta p,
\end{equation}
where $s$ is the fluid entropy per unit mass and $\delta$ stands for the variation of the thermodynamic
quantity during the time interval $\delta t$.
From mass conservation it follows
$\langle\delta\rho\rangle=0$, and, from the pressure homogeneity that
$\langle\delta p\rangle=\delta p$.

By averaging Eq.~(\ref{dr}) one obtains
\begin{equation}\label{dra}
    \langle\left({\partial\rho\over\partial s}\right)_p\delta s\rangle+
    \langle\left({\partial\rho\over\partial p}\right)_s\rangle\,\delta p=0.
\end{equation}
The use of appropriate thermodynamic relationships leads to
\begin{equation}\label{dp1}
  \delta p={\left\langle{\chi_T\over c_p}\,T\left({\partial p\over\partial T}\right)_\rho\,\rho\delta s\right\rangle
    \over\left\langle{\chi_T\over c_p}\,\rho c_v\right\rangle}.
\end{equation}
In order to use the second law of thermodynamics
\begin{equation}\label{law}
   \langle\rho\delta s\rangle={\delta Q\over v\bar T},
\end{equation}
where $\delta Q$ is the total change of the amount of heat of the fluid,
one needs to separate out the
averages of the form $\langle YZ\rangle$ in Eq.~(\ref{dp1}).
Under the assumption that
$Y$ (or $Z$) does not vary sharply over the fluid volume, the following approximation holds
(see Appendix~\ref{apB}):
\begin{equation}\label{YZ}
    \langle YZ\rangle\approx\langle Y\rangle\langle Z\rangle.
\end{equation}
Among the quantities that appear in  Eq.~(\ref{dp1}), only $\chi_T$ and $c_p$ vary sharply near the critical point
and could thus vary strongly across the fluid volume.
However, only their ratio, which  remains constant near
the critical point, enters Eq.~(\ref{dp1}). Hence, the average
of this ratio as well as the remainder averages of slowly varying quantities
can be separated.
By using the expression for the total heat change rate
\begin{equation}\label{q}
    {\delta Q\over \delta t}=\int_A k{\partial T\over \partial \vec n}\textrm{d} A,
\end{equation}
where the r.h.s. is simply the integrated heat flux supplied to the fluid through its boundary $A$ (with
$\vec n$ the external normal vector to it), one gets to the final expression
\begin{equation}\label{Ti}
{\textrm{d}\bar T\over  \textrm{d} t}={1\over\langle\rho\rangle vc_v}\int_A  k{\partial T\over
\partial \vec n}\textrm{d} A.
\end{equation}
In the fast calculation method, Eq.~(\ref{Ti}) plays the role of the pressure equation (\ref{pB}). Equation
(\ref{Ti}) is both substituted directly into Eq.~(\ref{gb}) and solved to get the temperature $\bar T$ using
the initial condition (\ref{initT}). The obtained value for $\bar T$ is used to solve Eq.~(\ref{tau}) and to
calculate all the fluid properties. Note that $\bar T$ should not be confused with $T$ from Eqs.~(\ref{q},
\ref{Ti}). The spatially varying fluid temperature $T$ has to be calculated with Eqs.~(\ref{psi},\ref{hb1}).

Substituting  $\bar T$ by $\langle T\rangle$, Eq.~(\ref{Ti}) coincides with the result of Onuki and Ferrell
\cite{Onuki} which was derived by a different way. Eq.~(\ref{Ti}), written in terms of $\langle T\rangle$,
was employed recently \cite{Onuki1, Onuki2} to simulate the gravitational convection in 2D by the finite
difference method. However, the finite difference numerical method is not the most efficient for the
computation of heat transfer problems.

\section{Hydrodynamic approach}\label{Anne}

Analytical analysis as well as direct simulations were carried out in previous works. Bailly and Zappoli
\cite{HBAIZAP} have developed a complete hydrodynamic theory of density relaxation after a temperature step
at the boundary of a cell filled with a nearly supercritical fluid in microgravity conditions. In
\cite{HBAIZAP}  they describe the different stages of the fluid relaxation towards its complete thermodynamic
equilibrium, covering the acoustic, Piston effect and heat diffusion time scale. The analytical approach
leans on the matched asymptotic expansions to solve the 1D Navier-Stokes equations for a viscous,
low-heat-diffusing, near-critical van der Waals fluid (see \cite{HZAPCAR} and \cite{HBAIZAP}). The DNS of the
Navier-Stokes equations were performed in 1D and 2D geometries. Some of them take into account gravity
effects, as for example, the interaction of a near-critical thermal plume with a thermostated boundary
\cite{HZAPAMI}. Numerical results are also available on thermo-vibrational mechanisms \cite{HJOU}.

To date the hydrodynamic approach has been solved for the classical, van der Waals, EOS. This EOS allows a
considerable reduction in computational time when compared to the restricted cubic EOS. However, it does not
provide a correct description of the real fluids. In particular, it fails to predict the critical exponents
for the divergence laws of the thermodynamic properties. In the present work we use a more realistic cubic
EOS to describe the fluid behavior in the near-critical region. Hereafter, we describe the methodology
suitable for a general EOS.

\subsection{Problem statement}

The hydrodynamic description leads to the following set of equations
\begin{eqnarray}
\frac{\mathrm{d}\rho}{\mathrm{d}t} + \rho\nabla\cdot\vec{u}=
0,\label{eq:cont}\\ \rho\frac{\mathrm{d}\vec{u}}{\mathrm{d}t}= -\nabla
p +\mu\nabla^2\vec{u},\label{eq:NS}\\
\rho\frac{\mathrm{d}e}{\mathrm{d}t}=\nabla\cdot(k\nabla
T)-p\nabla\cdot\vec{u}+\Phi, \label{eq:ener}
\end{eqnarray}
where $e$ is internal energy per unit mass, $\vec{u}=(u_1,u_2,u_3)$ is the fluid velocity at the point
$\vec{x}=(x_1,x_2,x_3)$,
$$\Phi=\mu\sum_{i,j}\left(\frac{\partial u_i}{\partial
x_j}\frac{\partial u_j}{\partial x_i}+\frac{\partial u_i}{\partial
x_j}\frac{\partial u_i}{\partial x_j}-\frac{2}{3}\frac{\partial
u_i}{\partial x_i}\frac{\partial u_j}{\partial x_j}\right)$$ is the
dissipation function due to the shear viscosity $\mu$ (the bulk
viscosity is neglected). The operator $\mathrm{d}/\mathrm{d}t$ is
defined as
\begin{equation}
    \frac{\mathrm{d}}{\mathrm{d}t}=\frac{\partial }{\partial t}+\vec{u}\cdot\nabla.
    \label{eq:d}
\end{equation}
The set of equations (\ref{eq:cont},\ref{eq:NS},\ref{eq:ener}) is closed by adding the EOS (\ref{EOS}).

\subsection{$c_v$-formulation}
In the DNS, the energy equation (\ref{eq:ener}) is re-written in terms
of temperature $T$.  This is achieved by expressing the internal
energy as a function of density and temperature so that one can make
use of the well known relation
\begin{equation}
    \frac{\mathrm{d}e}{\mathrm{d}t}=\frac{1}{\rho^2}\left[p-T\left(\frac{\partial p}{\partial T}\right)_{\rho}\right]
    \frac{\mathrm{d}\rho}{\mathrm{d}t}+c_{v}\frac{\mathrm{d}T}{\mathrm{d}t}.
    \label{eq:e}
\end{equation}
Then, by substituting the Eqs.~(\ref{eq:cont}) and (\ref{eq:ener}) into Eq.~(\ref{eq:e}) one obtains:
\begin{equation}
\rho c_v\frac{\mathrm{d}T}{\mathrm{d}t}=\nabla\cdot(k\nabla T)-T\left(\frac{\partial p}{\partial
T}\right)_{\rho}\nabla\cdot\vec{u}+\Phi.    \label{eq:efin}
\end{equation}
Note that Eq. (\ref{eq:efin}) involves $c_v$ and not $c_p$ as in the thermodynamic Eq. (\ref{hcond}). The
``$c_{v}$-formulation" is preferred to the ``$c_{p}$-formulation" because the much weaker near-critical
divergence of $c_v$ (in comparison to $c_p$) allows $c_{v}$ to be assumed constant.

The boundary conditions for the Navier-Stokes equations are $\vec{u}=0$ at the walls. The initial conditions
are given by $\vec{u}(t=0)=0$.  For the energy equation (\ref{eq:efin}) the boundary and initial conditions
are identical to those applied in the energy equation (\ref{hcond}) (cf. Sec.~\ref{meth}). The values of the
physical parameters used in the simulations are discussed in Appendix \ref{prop}.

\subsection{Acoustic filtering}\label{Acc}

Heat transfer in supercritical fluid involves three characteristic time scales \cite{HGARBON,Jphys}: the
acoustic time scale defined by $t_a=L/c_0$ (where $c_0$ is the sound velocity and $L$ is the cell size), the
diffusion time scale $t_D=L^2/D$ ($D$ being the thermal diffusivity) and the Piston Effect time scale defined
by $t_{PE}=L^2/[D(c_p/c_v-1)^2]$, with $t_a\ll t_{PE}<t_D$. The present study is mainly concerned with time
of the same order as the Piston effect time scale so that a fine description of the acoustic phenomena is not
needed. This suggests that one can filter out the acoustic motions of the set of
Eqs.~(\ref{EOS},\ref{eq:cont},\ref{eq:NS},\ref{eq:efin}) and retain only their integrated effects without
altering the physics of our problem. The removal of the acoustic motions is achieved by applying the acoustic
filtering method \cite{HPAO} which is broadly used in the computation of the low Mach number compressible
Navier-Stokes equations because it avoids numerical instabilities when time steps, $\Delta t>>t_a$, are used
in the simulations. The following presents the main points of the acoustic filtering method.

The equation of momentum is first rewritten by choosing the sound
velocity $c_{0}$ as the reference velocity scale
and $L/u_{0}$ as the reference time scale (here $u_{0}$
is the characteristic velocity of large scale fluid motions, in our
case $u_{0}=L/t_{PE}$). Using this time and velocity scale the
Mach number $\mathrm{Ma}=u_0/c_0$ appears in the
non-dimensional momentum equation as follows,
\begin{equation}\label{filt}
    \rho\left[\frac{\partial\vec{u}}{\partial t}+\mathrm{Ma}^{-1}(\vec{u}\cdot\nabla)\vec{u}\right]=
    -{\mathrm{Ma}^{-1} p_c\over {c_0^2}\rho_c}\nabla p+{1\over \mathrm{Re}}\nabla^2\vec{u},
\end{equation}
where $\mathrm{Re}=\rho_c u_0 L/\mu$ is the Reynolds number and the density and pressure are
non-dimensionalized by the critical density $\rho_{c}$ and critical pressure $p_{c}$ taken as the reference
values. For small Mach numbers, one can express the fluid variables as series of Ma,
\begin{eqnarray}
    \vec{u}=\mathrm{Ma}[\vec{u}^{(0)}+\mathrm{Ma}^{2}\vec{u}^{(1)}+o(\mathrm{Ma}^{2})],\label{devv}
    \\
    p=p^{(0)}+\mathrm{Ma}^{2}p^{(1)}+o(\mathrm{Ma}^{2}),\label{devp}
\end{eqnarray}
While $\vec{u}$ in the l.h.s. of Eq.~(\ref{devv}) is non-dimensionalized with $c_0$, the term in the square
brackets defines the velocity non-dimensionalized with $u_0$. This explains the factor Ma in
Eq.~(\ref{devv}). The density and temperature are expanded like $p$ in Eq.~(\ref{devp}). By substituting the
series (\ref{devv}, \ref{devp}) into Eq.~(\ref{filt}) and neglecting the terms of order $O(\mathrm{Ma})$, one
obtains $\nabla p^{(0)}=0$, which means that $p^{(0)}$ depends on time only. By retaining $O(\mathrm{Ma})$
terms in Eqs.~(\ref{eq:cont},\ref{eq:efin},\ref{filt}) and $O(1)$ terms in the EOS (\ref{EOS}), one obtains
the final (dimensional) form for the governing equations:
\begin{eqnarray}
\frac{\mathrm{d}\rho^{(0)}}{\mathrm{d}t} =-\rho^{(0)}\nabla\cdot\vec{u}^{(0)},\label{contf}\\
\rho^{(0)}\frac{\mathrm{d}\vec{u}^{(0)}}{\mathrm{d}t}= -\nabla p^{(1)}
+\mu\nabla^2\vec{u}^{(0)},\label{NSf}\\
\rho^{(0)} c_v^{(0)}\frac{\mathrm{d}T^{(0)}}{\mathrm{d}t}=-T^{(0)}\left(\frac{\partial p}{\partial
T}\right)_{\rho}\nabla\cdot
\vec{u}^{(0)}+\nonumber\\\nabla\cdot(k\nabla T^{(0)}), \label{enerf}   \\
\Lambda(p^{(0)},\rho^{(0)},T^{(0)})=0, \label{EOSf}
\end{eqnarray}
where $(p_c/c_0^2\rho_c)p^{(1)}$ is replaced by $p^{(1)}$ for the sake of compactness. The pressure term
$p^{(1)}$ has to be interpreted as the dynamic pressure that makes the velocity field satisfy the continuity
equation (\ref{contf}). This term reflects the contribution of the acoustic waves averaged over several wave
periods to the total pressure field. One notes that the velocity scale $c_0$ is not present any more in
Eqs.~(\ref{contf}-\ref{EOSf}) which was the main purpose of the acoustic filtering.

The assessment of $p^{(0)}$ requires one more equation to close the set (\ref{contf}-\ref{EOSf}). This
additional equation expresses the mass conservation:
\begin{equation}
    {1\over v}\int_{v}\rho^{(0)} dv=\langle\rho\rangle,
    \label{conserv}
\end{equation}
where $\langle\rho\rangle$ is a known constant.

In the following, the superscript $(0)$ is dropped to conform to the
notation of Sec.~\ref{meth}.

\subsection{Numerical procedure}

For the time integration, the first order Euler scheme is
used. Equations ~(\ref{contf}-\ref{enerf}) are solved by the
iterative SIMPLER algorithm and by
applying the Finite Volume
Method (FVM, see Appendix~\ref{FVM}) on each grid cell of the
1D cell. Near the walls the mesh is refined
to properly resolve the very
thin thermal boundary layers.

In the present work the thermodynamics variables are determined using
the parametric EOS \cite{Seng}. This uses two parameters $r$ and
$\theta$ which both depend on temperature $T$ and density $\rho$.
Therefore, one needs to solve two equations
\begin{equation}\label{EOS3} \begin{array}{cr}
  &\Lambda_1(r_i,\theta_i,T_i)=0\\
  &\Lambda_2(r_i,\theta_i,p)=0 \\
\end{array}
\end{equation}
instead of one Eq.~(\ref{EOS}) for each volume element $i$ and time step.

The whole numerical procedure consists in solving by the
Newton-Raphson method, at each time step, a set of equations
that includes Eqs.~(\ref{EOS3}), written for each volume element and
Eq.~(\ref{conserv}).  This makes a system of $2N+1$
equations to resolve, $N$ being the
total number of the volume elements. The local temperature $T_i$
is given by the resolution of
Eqs.~(\ref{contf}-\ref{enerf}) at each iteration
of the SIMPLER algorithm for each time step, as described
in Appendix \ref{FVM}. For each value $T_i$ the $2N+1$
$(r_i,\theta_i,p)$ variables are computed via the system (\ref{EOS3}).

\section{Results and discussion}\label{Results}
\begin{figure}[hbt]
  \begin{center}
  \includegraphics[height=6cm]{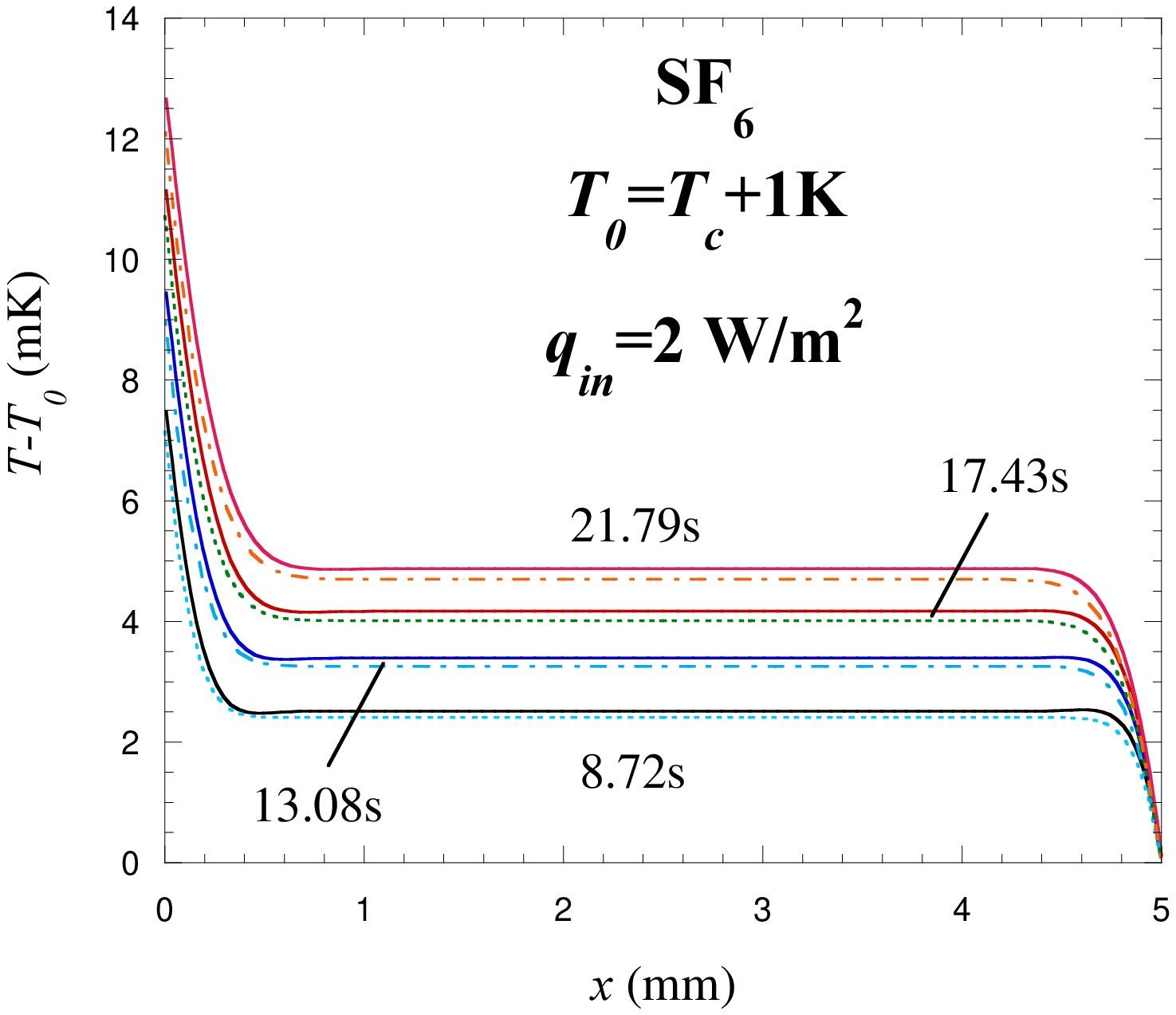}\\
  \small{(a)}\\
  \includegraphics[height=6cm]{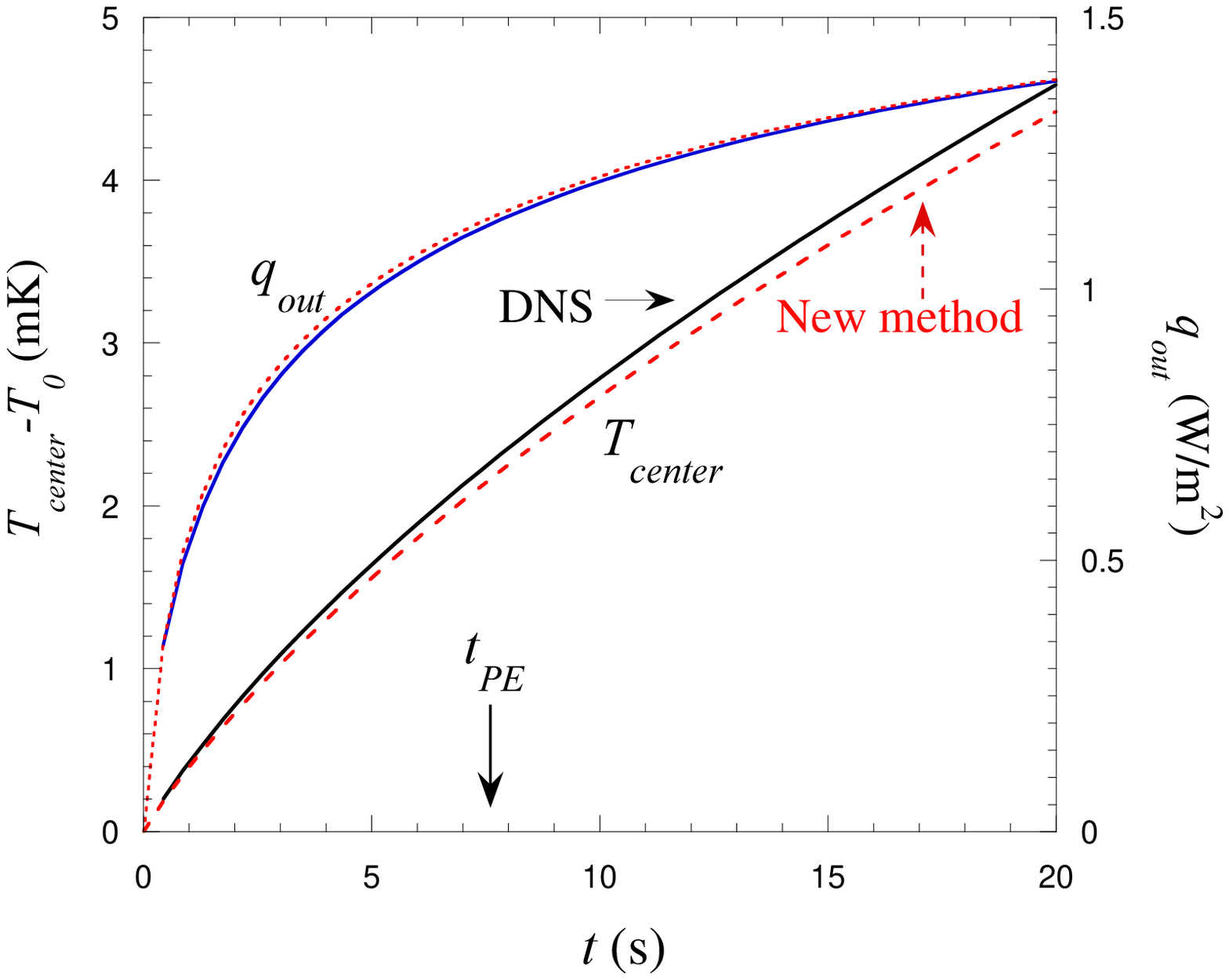}\\
  \small{(b)}
  \end{center}
\caption{Comparison of two approaches for SF$_6$ at 1K above $T_c$
(reduced temperature $3.1\cdot 10^{-3}$), $q_{in}=2$ W/m$^2$ and
$\langle\rho\rangle=\rho_c$. Solid curves are the DNS results and the
dotted curves are the new method results. (a) Spatial variation of the
temperature at different times. (b) Time evolution of the temperature
at the cell center and of the flux at the exit of the cell. The value
of $t_{PE}=7.73$ s obtained with our EOS is shown by an
arrow.}\label{SF6}
\end{figure}

A brief analysis comparing the $c_p$ and $c_v$ formulations
(\ref{hcond}) and (\ref{enerf}) of the energy equation allows us to
gain more insight into the relation between the two
approaches. Formally, Eq.(\ref{hcond}) and Eq.(\ref{enerf}) become
equivalent if the advection term
\begin{equation}\label{conv}(\vec{u}\cdot\nabla) T
\end{equation} is added to l.h.s. of
Eq.~(\ref{hcond}). However, the equivalence of the two forms under
which the pressure work appears (see second term of the l.h.s of
Eq.(\ref{hcond}) and Eq.(\ref{enerf})) is not trivial and deserves
to be detailed.  At the early stage of the heating ($t<t_{PE}$) the
velocity at the front of the cold boundary layer being very small, the
velocity can be assumed to
decrease linearly in the bulk cell as
$\partial u/\partial x \simeq -u_{max}/L$, where $u_{max}$ is the
maximum velocity located at the front of the hot boundary layer and
$x$ the distance from the hot wall to the cold wall.  The rate of temperature
increase due to the pressure contribution in Eq.~(\ref{enerf}) can
thus be written as follows,
\begin{equation}\label{peterm}
-\frac{T}{\rho c_{v}}\left(\frac{\partial p}{\partial T}\right)_{\rho}\nabla\cdot \vec{u}=\frac{T}{\rho
c_{v}}\left(\frac{\partial p}{\partial T}\right)_{\rho}\frac{u_{max}}{L}
\end{equation}
By using the expression \cite{HGARBON},
\begin{equation}\label{umax} u_{max}=\frac{1}{T}\left(\frac{\partial
T}{\partial p}\right)_\rho\frac{\delta Q}{A \delta t },
\end{equation}
and Eqs.~(\ref{q},\ref{Ti}) one concludes that the term (\ref{peterm}) is equivalent to (\ref{gb}) near the
critical point where $c_p\gg c_v$. One can note that in the hydrodynamic approach, the pressure work is
directly related to the mass transfer from the hot boundary layer to the bulk fluid via the gradient
velocity. It is then very important to asses properly the effect of the velocity field in order to compare
the fast calculation and hydrodynamic methods.  The above analysis has shown that the expressions of the
pressure work is equivalent for both methods. Hence, the remaining potential interaction between the velocity
and energy fields can manifest itself only through the advection term (\ref{conv}). This term is only
relevant when, at the same spot of the fluid, both the fluid velocity and the temperature gradient are large.
At the small times, $t<t_{PE}$, the temperature gradients are confined very near the wall where the velocity
remains small \cite{Jphys}. Later on, the velocity maximum shifts to the center of the cell where the
temperature gradient is small. At very large times, $t>t_D$, the Piston effect is not efficient and the
velocity tends to zero. We thus do not expect a strong influence of the advection effects on the temperature
field. This will be confirmed by the results presented below.  Note that the advection term cannot be
neglected when the flux distribution over the heater surface is highly inhomogeneous. Hot jets \cite{jets}
can be generated in this case.

\begin{figure}[bt]
  \begin{center}
  \includegraphics[height=6cm]{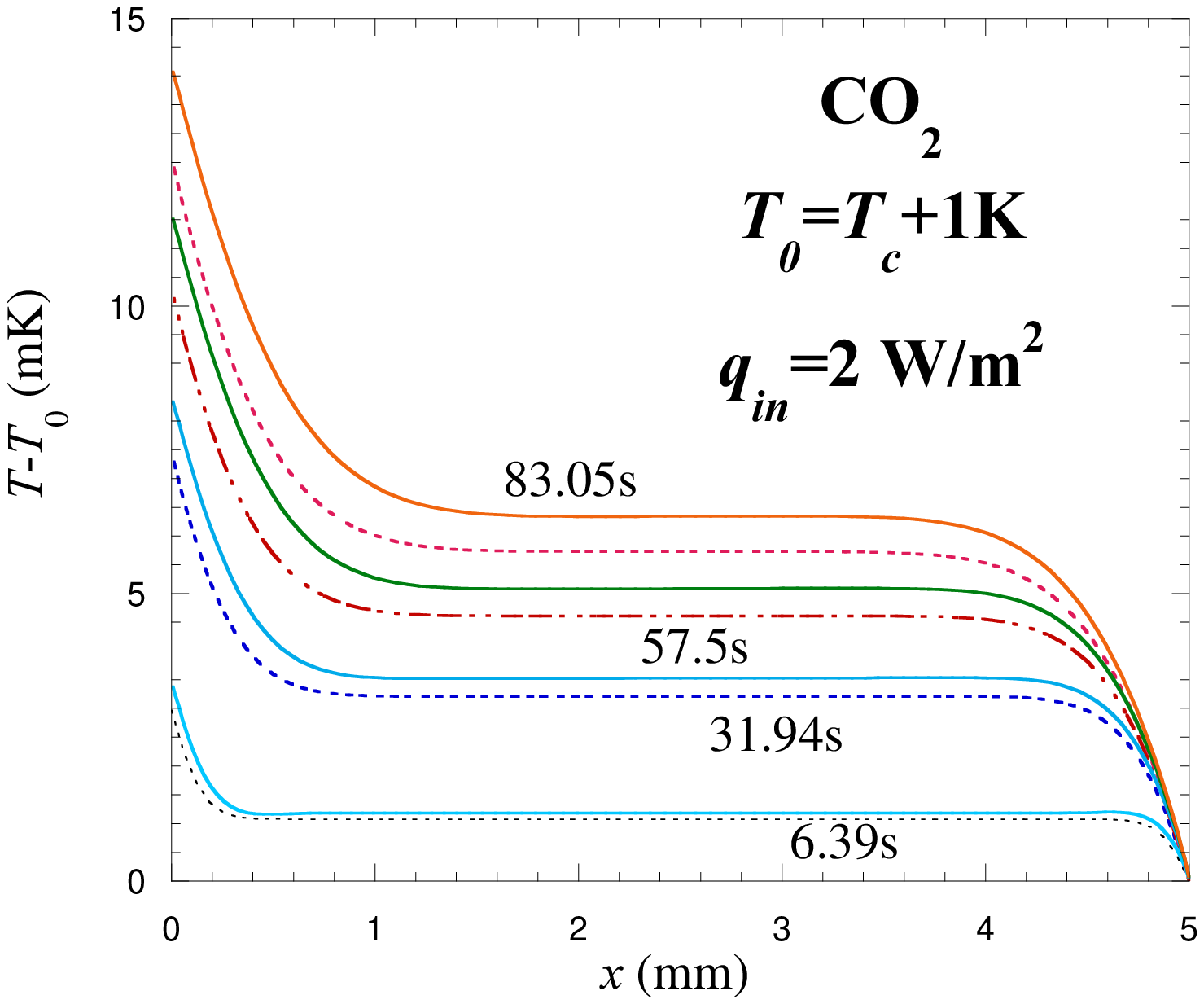}\\
  \small{(a)}\\
  \includegraphics[height=6cm]{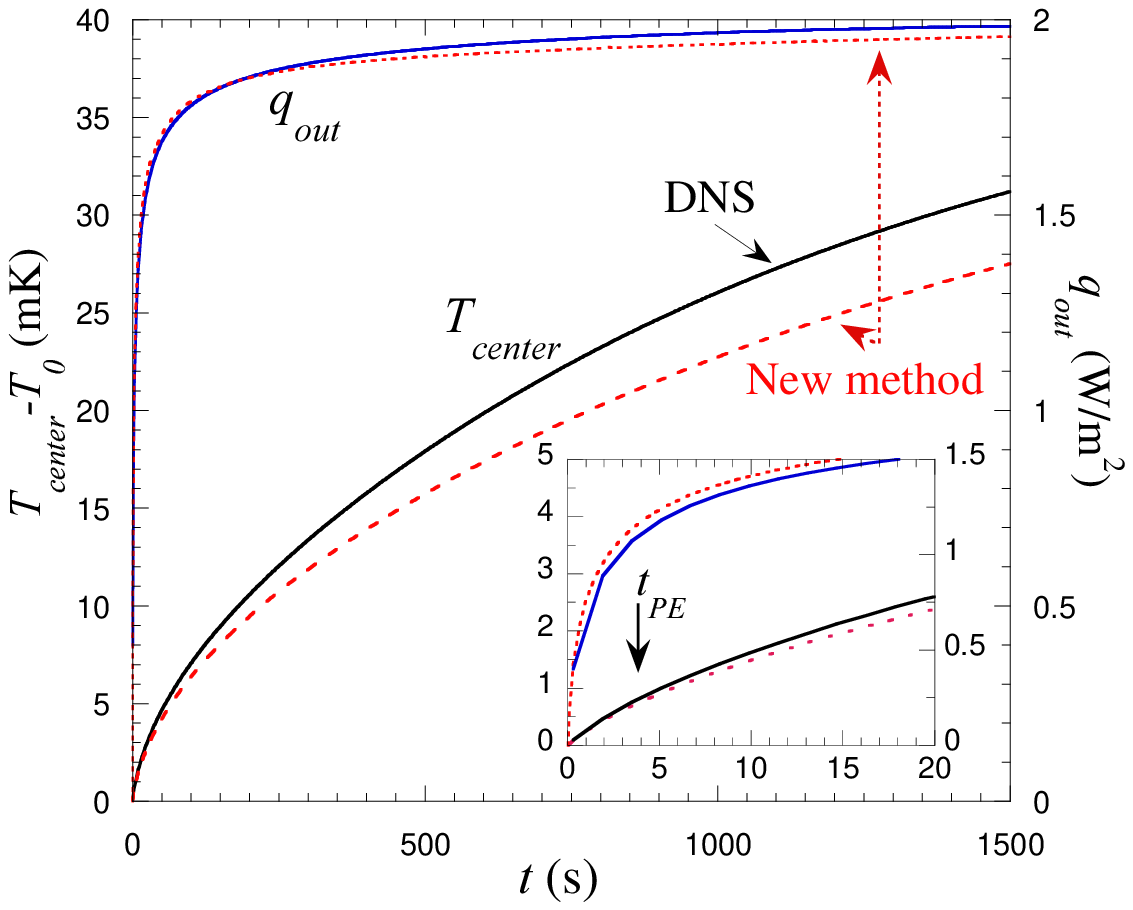}\\
  \small{(b)}
  \end{center}
\caption{Comparison of the two approaches for CO$_2$ at 1K above $T_c$
(reduced temperature $3.3\cdot 10^{-3}$), $q_{in}=2$ W/m$^2$ and
$\langle\rho\rangle=\rho_c$. Solid curves are the DNS results and the
dotted curves are the new method results. (a) Spatial variation of the
temperature at different times. (b) Time evolution of the temperature
at the cell center and of the flux at the exit of the cell. The value
of $t_{PE}=3.45$ s obtained with our EOS is shown by an arrow in the
insert that presents the short time evolution.}\label{CO21}
\end{figure}

The calculations have been performed for two fluids, CO$_2$ and SF$_6$, confined in a cell of length $L=5$mm.
The initial temperatures 1K and 5K above the critical point have been considered for CO$_2$. The computations
related to SF$_6$ concern only the initial temperature 1K above the critical point. The cell boundary
situated at $x=0$ has been submitted to the constant heat flux $q_{in}=2$W/m$^{2}$ (for $T_{0}=T_c +1$K) and
$q_{in}=9.5$W/m$^{2}$ (for $T_{0}=T_c +5$K) and the opposite boundary has been maintained at the constant
temperature $T(x=L)=T_0$.

The time evolution of the temperature profiles and the temperature at the cell center $T_{center}=T(x=L/2)$
as well as the heat flux $q_{out}=-k{\partial T(x=L)\over\partial x}$ are compared and analyzed. In the case
of CO$_2$, the time evolution covers not only the Piston effect time scale $t_{PE}$ but also the large
diffusion time scale.
\begin{figure}[hbt]
  \begin{center}
  \includegraphics[height=6cm]{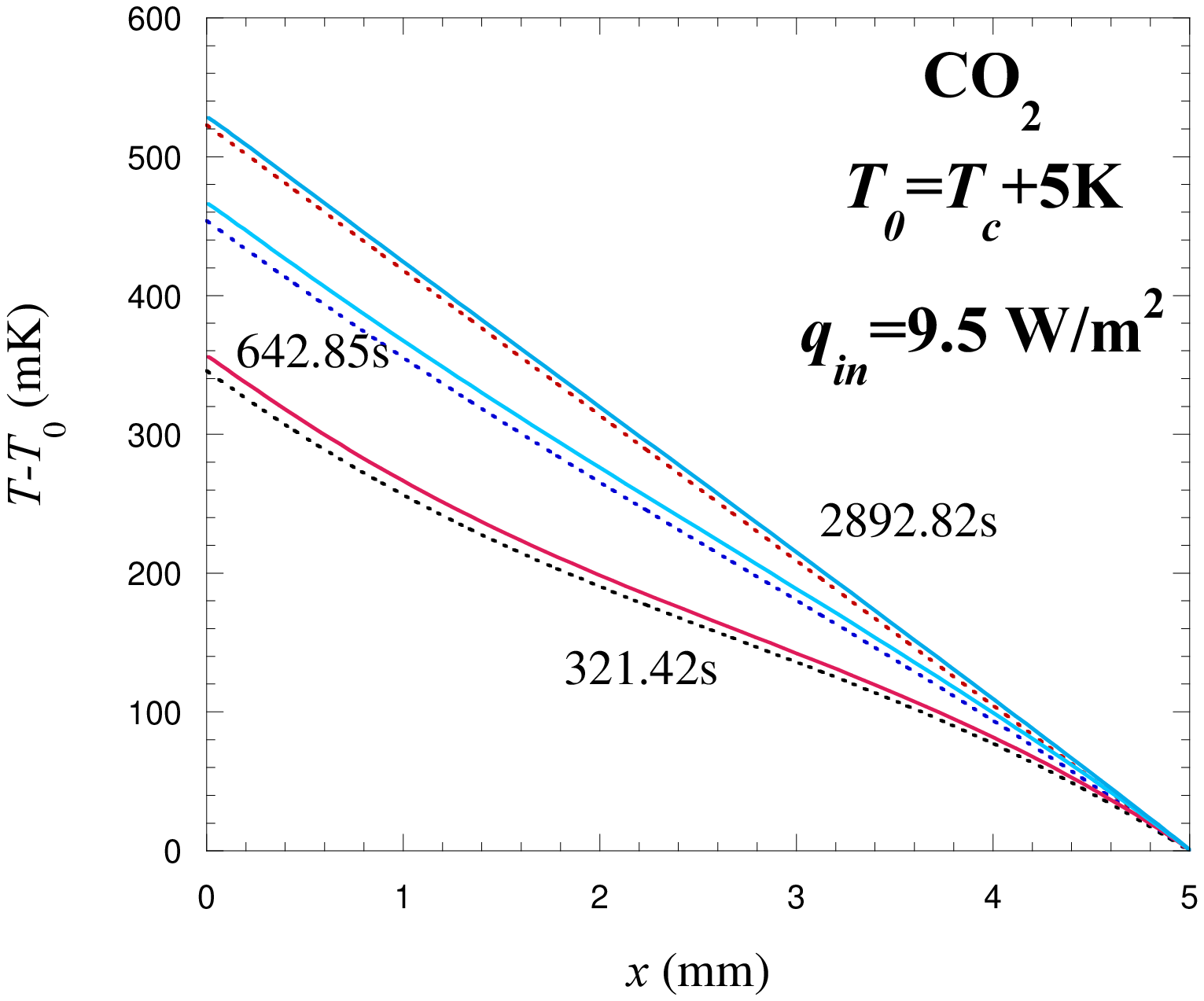}\\
  \small{(a)}\\
  \includegraphics[height=6cm]{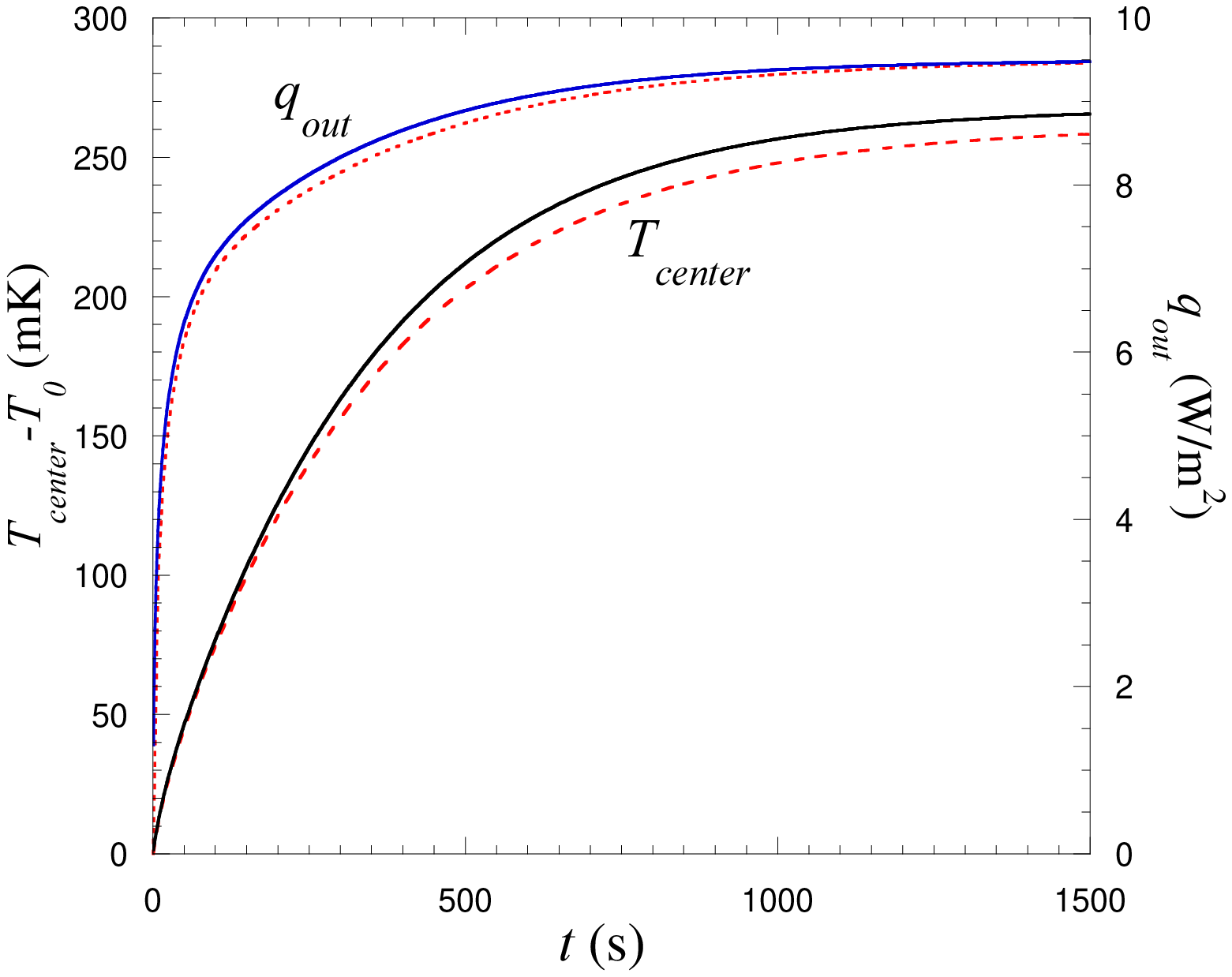}\\
  \small{(b)}
  \end{center}
\caption{Comparison of two approaches for CO$_2$ at 5K above the
critical temperature (reduced temperature $1.6\cdot 10^{-2}$), $q_{in}
=9.5$ W/m$^2$ and $\langle\rho\rangle=\rho_c$. Solid curves are the
DNS results and the dotted curves are the new method
results. According to our EOS, $t_{PE}=25.67$ s. (a) Spatial variation
of the temperature at different times. (b) Time evolution of the
temperature at the cell center and of the flux at the exit of the
cell.}\label{CO25}
\end{figure}

The set of Figs.~\ref{SF6}, \ref{CO21}, and \ref{CO25} exhibit a very good qualitative agreement between the
DNS and the fast calculation results. The thin boundary layers and the homogeneous enhancement of the
temperature in the center cell are very well predicted.  The quantitative comparison sets out two behaviors.
On one hand, the flux $q_{out}$ appears to fit very well with the DNS over the full time evolution, at the
time scale $t_{PE}$ as well at the time scale $t_D$, see Figs.~\ref{CO21}b, \ref{CO25}b. On the other hand,
the temperature at the cell center $T_{center}$ tends to be lower than the DNS data. This discrepancy
increases with time and is larger when the temperature is closer to the critical temperature (see
Fig.~\ref{CO21}b and Fig.~\ref{CO25}b). Both behaviors can be explained by considering how the thermal
conductivity $k$ is estimated in each method. In the hydrodynamic approach $k$ is determined locally, whereas
the fast calculation uses the spatial average value of $k$. Thus, keeping in mind that the thermal
conductivity diverges when approaching the critical point, the increment in temperature near the heating
surface tends to be smaller in the new method than in the DNS (see Fig. \ref{CO21}a).  At the opposite
surface the temperature is fixed at the initial temperature and is closest to the bulk temperature so that
the effect of averaging $k$ is less influent in this region. One can note that the thermal diffusivity can be
computed locally in the fast calculation method by applying the Kirchhoff substitution of the dependent
variable $\psi$ (defined by Eq.~(\ref{psi})) by $\phi=\int_0^\psi k(\psi)\textrm{d}\psi$.

A physical interpretation of the temperature $\bar T$, which was
formally introduced in Sec. \ref{meth}, can now be given. Indeed,
since in the fluid bulk (i.e., arround the center of the cell)
${\partial T\over\partial x}=0$ at $t<t_{PE}$, according to
Eq.~(\ref{hcond}) we have $\partial T_{center}/\partial t=\bar g(\bar T)$. As
near the critical point $c_p\gg c_v$, Eq.~(\ref{gb}) provides $\bar
g(\bar T)\approx \partial \bar T/\partial t$.  Finally, one can
conclude that $\bar T\approx T_{center}$. In other words, $\bar T$ can
be considered as the bulk temperature.

Aa a further remark, we note that for 1D the Eq.~(\ref{hpsi}) could have been solved analytically by series
expansion. Nevertheless, we prefer the use of the BEM for its generality and its possible extension to higher
dimensions.  We note that in 2D and 3D the BEM remains advantageous in resolving linearized problems when
compared to other numerical methods. Its success is based on several factors. One of them is its numerical
stability: the numerical solution of the integral equations is much more stable than that of the differential
equations and allows the use of larger time steps. Another advantage consists in the possibility of
determining analytically the BEM coefficients, Eqs.~(\ref{GBEM}, \ref{HBEM}). For 2D configurations, the
diagonal coefficients $G_{FF}$ and $H_{FF}$ (which have the largest absolute value and thus are the most
relevant) can be calculated analytically. The semi-analytical integration can be used for the remaining
coefficients \cite{IJHMT,BEM24}.

\section{Conclusions}\label{Concl}

In this work we propose a thermodynamic method for describing the heat transfer in supercritical fluids in
absence of gravity effects.  The method has been compared with the solution of the full hydrodynamic
equations, showing an excellent agreement.  In general, a thermodynamic approach leans on the possibility of
expressing the pressure work independently of the velocity field.  If so, the transfer of momentum does not
need to be considered, allowing a large reduction in computational time. As an example, in calculations
carried out for CO$_2$ and SF$_6$, the present thermodynamic method within minutes provided the complete
evolution of the heat transfer process, while the direct numerical simulation of the full hydrodynamic
equations required weeks of CPU time.

Compared with previous thermodynamic methods \cite{Bouk},
the fast calculation method presented here does not require the
evaluation of the variables at each cell of the computation domain.
This fact ensures a much better performance.
Moreover, the proposed method offers the possibility
to  explicitly include the thermal behaviour
of the material vessel containing the fluid by taking into account the
heat conduction along the solid walls, see Ref. \cite{Caloduc}.

The direct numerical simulation of the flow has been used to analyze the validity of the method proposed
here. The accuracy of the latter approach is explained by the fact that the advection of energy remains
negligible.

For the sake of completeness,
we have also presented a detailed description of the hydrodynamic
approach. While it has been used for about a decade, some parts of its
description for a general equation of state are either
dispersed over many literature sources or not published at all in the
accessible literature.

Concerning the future development of the present research, we
plan to extend the fast calculation
method to two- and three-dimensional problems.
Finally, we intend to use this method to investigate the
heat transfer in two-phase fluids.

\begin{acknowledgments}
This work was partially supported by the CNES. We thank Carole Lecoutre for her help with the code launching.
A.~D. acknowledges Jalil Ouazzani for the helpful discussions on the numerical method used in the
hydrodynamic approach.
\end{acknowledgments}

\appendix

\section{BEM for the diffusion equation}\label{appA}

In this Appendix we use the traditional notation, so that $D$ and $t$ correspond to $D_d$ and $\tau$ of
Eq.~(\ref{hb1}). It can be shown \cite{BEM} that the linear diffusion problem
\begin{equation}\label{hpsi}
\begin{array}{c}\displaystyle
  {\partial \psi\over\partial t}=D\nabla^2\psi \\
  \left.\psi\right|_{t=0}=0 \\
\end{array}
\end{equation}
with the constant thermal diffusion coefficient $D$ is equivalent to the boundary integral equation
\begin{eqnarray}
D\int\limits_{0}^{t}{\rm d}t'\int\limits_A\Biggl[G(\vec{x}-\vec{x'},t-t') {\partial_{x'} \psi(\vec{x'},t')\over\partial\vec{n}}- \nonumber\\
\left.\psi(\vec{x'},t'){\partial_{x'} G(\vec{x}-\vec{x'},t-t')\over\partial\vec{n}} \right]{\rm
d}_{x'}A={1\over 2}\psi(\vec{x},t).\label{Ieq}
\end{eqnarray}
The integration is performed over the surface $A$ of the fluid volume $v$, $\vec{x}\in A$. The outward unit
normal to $A$ is $\vec{n}$. The Green function $G$ for the infinite space for the equation adjoint to
Eq.~(\ref{hpsi}) reads
\begin{equation}\label{G}
G(\vec{x},t)=(4\pi Dt)^{-d/2}\exp\left(-{|\vec{x}|^2\over 4D t}\right),\end{equation} where $d$ is the
spatial dimensionality of the problem (\ref{hpsi}).

Only the 1D case with the space variable $x\in(0,L)$ is considered below, the 2D counterpart being described
elsewhere \cite{IJHMT,BEM24}. For $d=1$, $A$ degenerates into two points, and Eq.~(\ref{Ieq}) reduces to two
equations
\begin{eqnarray}
2D\int\limits_{0}^{t}{\rm d}t'\Biggl[G(x-x',t-t') {\partial\psi\over\partial x'}- \nonumber\\
\left.\left.\psi(x',t'){\partial G(x-x',t-t')\over\partial x'} \right]\right|_{x'=0}^{x'=L}=\psi(x,t)\quad
\label{I1D}
\end{eqnarray}
written for $x=0,L$. A variety of numerical methods can be applied to solve Eqs.~(\ref{I1D}). The simplest
way is to present the integral over $(0,t)$ as a sum of the integrals over $(t_{f-1},t_f)$, $f=1..F$ with
$t_0=0$ and $t_F=t$ and assume a constant values $\psi_f=\psi(t_f)$ and $\psi'_f=\partial\psi(t_f)/\partial
x$ over each of these intervals. Eqs.~(\ref{I1D}) will reduce then to the set of $2F_m$ ($t_{F_m}$ is the
maximum desired calculation time) linear equations
\begin{eqnarray}\label{set}
\sum\limits_{f=1}^F[\psi_{f}(0)H_{Ff}(x)-\psi_{f}(L)H_{Ff}(x-L)-\nonumber\\
\psi'_{f}(0)G_{Ff}(x)+\psi'_{f}(L)G_{Ff}(x-L)]=\nonumber\\\psi_{Fi}/2,\quad x=0,L;\; F=1,F_m
\end{eqnarray}
for $4F_m$ variables $\psi_f$, $\psi'_f(0,L)$, $2F_m$ of them being defined by the boundary conditions. The
coefficients in these equations can be calculated analytically:
\begin{eqnarray}\label{GBEM}
 G_{Ff}(x)=D\int_{t_{f-1}}^{t_f}G(x,t_F-t'){\rm d}t'=\nonumber\\
 {|x|\over 2}\left.\left[\mbox{erfc}(\sqrt{u})-{\exp(-u)\over\sqrt{u\pi}}\right]
 \right|_{x^2\over4D(t_F-t_{f-1})}^{x^2\over4D(t_F-t_f)}
\end{eqnarray}
and
\begin{eqnarray}\label{HBEM}
 H_{Ff}(x)=-D\int_{t_{f-1}}^{t_f}{\partial G(x,t_F-t')\over\partial x}{\rm d}t'=\nonumber\\
 -\left.{\mbox{sign}(x)\over 2}\mbox{erfc}(\sqrt{u})
 \right|_{x^2\over4D(t_F-t_{f-1})}^{x^2\over4D(t_F-t_f)},
\end{eqnarray}
where
 $$\mbox{erfc}(x)=\int_x^\infty\exp(-u^2)\textrm{d}u,$$
is the complementary error function and
 $$\mbox{sign}(x)=\left\{\begin{array}{rc}
   1, & x>0, \\
    -1, & x<0.
 \end{array}\right.$$
 One needs to mention that the case $x=0$ is special: $H_{Ff}(0)=0$ for all $f$ and $F$ and
 $$G_{Ff}(0)=\sqrt{D_0/\pi}(\sqrt{t_F-t_{f-1}}-\sqrt{t_F-t_f}\,).$$

 The set of linear equations (\ref{set}) can be solved by any appropriate method, e.g. by the Gauss elimination.

\section{The fluid properties}\label{prop}

The thermal conductivity $k$ is deduced from the thermal diffusivity
\begin{equation}
    D=D_{1}{\left(\frac{T-T_{c}}{T_{c}}\right)}^{\varphi_{1}}+
    D_{2}{\left(\frac{T-T_{c}}{T_{c}}\right)}^{\varphi_{2}}
    \label{eq:diffus}
    \end{equation}
and the constant pressure specific heat at critical density $\rho_c$, $k=D\rho_{c}c_{p}|_{\rho_c}$.
Coefficients values for CO$_2$ are:

$D_{1}=5.89184\times 10^{-8}$ m$^{2}$/s, $D_{2}=7.98068\times 10^{-7}$ m$^{2}$/s, $\varphi_{1}=0.67$ and
$\varphi_{2}=1.24$.

Coefficients values for SF$_6$ are:

$D_{1}=6.457\times 10^{-7}$ m$^{2}$/s, $D_{2}=0$, $\varphi_{1}=0.877$ and $\varphi_{2}=0$.

The specific heat at constant pressure is calculated by using the thermodynamic relationship
      \begin{equation}
     c_{p}=c_{v}+T\left(\frac{\partial p}{\partial T}\right)_\rho^{2}\chi_T
     \label{eq:Cp}
 \end{equation}
The isothermal compressibility coefficient $\chi_{T}$ and the specific heat at constant volume are given by
the restricted cubic model \cite{Seng}. For the reference hydrodynamic DNS we used a constant $c_v$ value
calculated for the initial value of temperature and density. We used a constant value for the viscosity
$\mu$: $3.74\cdot 10^{-5}$ Pa$\cdot$s for SF$_6$ and $3.45\cdot 10^{-5}$ Pa$\cdot$s for CO$_2$.

\section{}\label{apB}

According to the integral theorem about the mean value \cite{Korn}, there is always a point $\vec{x}_m\in v$
so that
\begin{equation}\label{eaB} \int_v
Y(\vec{x})Z(\vec{x})\mathrm{d}\vec{x}=Y(\vec{x}_m)\int_v Z(\vec{x})\mathrm{d}\vec{x}
\end{equation}
if the functions $Y,Z$ are continuous. When the spatial variation of $Y$ in $v$ is small, $\langle
Y\rangle\approx Y(\vec{x}_m)$ and Eq.~\ref{YZ} stems from Eq.~(\ref{eaB}).

\section{Application of the Finite Volume Method (FVM) and SIMPLER algorithm}\label{FVM}

According to the FVM, the calculation domain is divided into a number of non-overlapping control volumes so
that there is one control volume surrounding each grid point. The differential equations are integrated over
each control volume. The attractive feature of this method is that the integral balance of mass, momentum,
and energy is exactly satisfied over any control volume (called below the cell for the sake of brevity), and
thus over the whole calculation domain. The integral formulation is also more robust than the finite
difference method for problems which present strong variations of properties observed in a near-critical
fluid \cite{HZAPDUR,HAMIOUA}. The equations are resolved on a staggered grid. This means that the velocity is
computed at the points that lie on the faces of the cell while the scalar variables (pressure, density and
temperature) are computed at the center of the cell. This choice is made to avoid pressure
oscillations in the computations \cite{HPAT}. For the time discretization, the first order Euler scheme is
used. For the sake of simplicity and clarity we present the finite volume method for the 1D generalized
transport equation for a variable $Y$ (where $Y$ can be substituted by either $u$ or $T$)
\begin{equation}
\frac{\partial \rho Y}{\partial t} +\frac{\partial \rho u Y}{\partial x} = \frac{\partial}{\partial
x}\left(\Gamma\frac{\partial Y} {\partial x}\right) + S \label{eq:transportphi}
\end{equation}
where $\Gamma$ denotes the generalized diffusion coefficient and $S$ the generalized source term
(volume forces). Integrated
over the $i$th cell of the length $\delta x$, Eq.~(\ref{eq:transportphi}) takes the form
\begin{equation}
\frac{\rho_{P} Y_{P}-\rho_{P}^p Y_{P}^p}{\Delta t} \Delta x +J_{e}-J_{w}= S_{P} \Delta x
\label{eq:transportphiint}
\end{equation}
where the superscript $p$ denotes the value on the previous time step, the subscript $P$ represents the
center of the cell, the subscripts $e$ and $w$ its ``east" and ``west" face respectively. The calculation of
the flux
\begin{equation}
J=\rho u Y -\Gamma \frac{\partial Y}{\partial x}
\end{equation}
on the faces requires the knowledge of $Y$ and $\rho$ at the \emph{centers} of two neighboring ``East" and
``West" cells denoted by the \emph{capital} letters E and W. Their values at the faces can be found by linear
interpolation between their values at the centers, e.g. $Y_e=0.5(Y_P+Y_E)$ if the nodes are equidistant.

The continuity equation integrated on the control volume is given by:
\begin{equation}
\frac{\rho_{P}-\rho_{P}^p}{\Delta t}\Delta x +F_{e}-F_{w}=0 \label{eq:continuityint}
\end{equation}
with $F=\rho u$.
When multiplying Eq.~(\ref{eq:continuityint}) by $Y_{P}$ and subtracting the result from
Eq.~(\ref{eq:transportphiint}), one obtains the equation
\begin{equation}
\frac{\rho_{P}^p \Delta x}{\Delta t}(Y_{P}-Y_{P}^p)
+(J_{e}-Y_{P}F_{e})-(J_{w}-Y_{P}F_{w})=S_{P}\Delta x
\end{equation}
that can be rewritten in the following form
\begin{equation}
a_{P}Y_{P}=a_{W}Y_{W}+a_{E}Y_{E}+b \label{eq:systemtridiag}
\end{equation}
The tridiagonal set of linear equations (\ref{eq:systemtridiag}) with respect to $Y_P$ is solved by the
Thomas algorithm \cite{HAND}. The stencil coefficients $a_{P}$, $a_{W}$ et $a_{E}$ depend on the
discretization scheme. Their general expression is
\begin{equation}\label{eq:stencil}\begin{array}{l}
  a_{W}=B_{w}A_{w} +\max(-F_{w},0), \\
  a_{E}=B_{e}A_{e} +\max(F_{e},0), \\
  a_{P}=a_{W}+a_{E}+\rho_{P}^p \Delta x/\Delta t, \\
  b=S_{P} \Delta x + \rho_{P}^pY_{P}^p \Delta x/\Delta t,  \\
\end{array}
\end{equation}
where  $B=\Gamma/\Delta x$. We use the ``power law scheme" \cite{HPAT} that requires
$$ A_i=\max\left[0,\left(1-\frac{0.1|F_i|}{B_i}\right)^{5}\right],\quad i=e,w.$$

The set (\ref{eq:systemtridiag}) should be written and solved both for the velocity and the temperature.
While the above scheme can be directly applied for the temperature case, the coupling of the velocity and the
pressure $p^{(1)}$ (which is defined implicitly by the continuity equation) requires a special treatment for
the velocity equation as described below.

The non-dimensionalized and discretized Navier-Stokes equation (\ref{NSf})
\begin{equation}
a_{e}u_{e}=\sum a_{nb}u_{nb}+(p^{(1)}_{P}-p^{(1)}_{E})+b, \label{eq:discrmomentum}
\end{equation}
where the subscript $nb$ denotes the neighbors of the point $e$, can be solved
only when the pressure field is given. Unless the correct pressure field is employed, the resulting velocity
field will not satisfy the continuity equation. We use the iterative SIMPLER algorithm \cite{HPAT} to couple
the velocity and the pressure fields. This algorithm is based on successive corrections
 of the velocity field and pressure field at a given time step. The velocity and
pressure variables are decomposed as follows:
\begin{equation}
\begin{array}{l}
  u=u^{*}+u', \\
  p^{(1)}=p^{(1)*}+p^{(1)\prime},
\end{array}
\end{equation}
where the asterisk denotes the guesses and prime the corrections. The steps of the SIMPLER algorithm are the
following:
\begin{enumerate}
    \item Start with a guessed velocity field.
    \item \label{st2}A pseudo-velocity $\widehat{u}$ (without taking into account
the pressure gradient) is first computed and is defined as
\begin{equation}
\widehat{u}_{e}=\frac{\sum a_{nb}u_{nb} +b}{a_{e}}
\end{equation}
where $u_{nb}$ represents the neighbor velocities. $\widehat{u}$ satisfies
\begin{equation}\label{uhat}
u_{e}=\widehat{u}_{e}+\frac{p^{(1)*}_{P}-p^{(1)*}_{E}}{a_{e}}.
\end{equation}
    \item Compute the pressure $p^{(1)*}$ whose equation is deduced by applying the
divergence operator to Eq.~(\ref{uhat}) and using the continuity equation (\ref{eq:continuityint}):
\begin{eqnarray}
\left(\frac{\rho_{w}}{a_{w}} + \frac{\rho_{e}}{a_{e}}\right)p^{(1)*}_{P}
=\frac{\rho_{w}}{a_{w}}p^{(1)*}_{W}+\frac{\rho_{e}}{a_{e}}p^{(1)*}_{E}
+\nonumber\\
\frac{\rho_{P}^p-\rho_{P}}{\Delta t} \Delta x
-\rho_{e}\widehat{u}_{e}+\rho_{w}\widehat{u}_{w}.\label{eq:pressimpler}
\end{eqnarray}
    \item Solve Eq.~(\ref{eq:discrmomentum}) with $p^{(1)*}$ used for $p^{(1)}$ and thus obtaining $u^{*}$.
    \item Compute $p^{(1)\prime}$ whose equation is obtained analogously to Eq.~(\ref{eq:pressimpler}) from
\begin{equation}
u_{e}=u^{*}_{e}+\frac{(p^{(1)\prime}_{P}-p^{(1)\prime}_{E})}{a_{e}} .\label{eq:corvel}
\end{equation}
It takes the form
\begin{eqnarray}
\left(\frac{\rho_{w}}{a_{w}} + \frac{\rho_{e}}{a_{e}}\right)p^{(1)\prime}_{P}
=\frac{\rho_{w}}{a_{w}}p^{(1)\prime}_{W}+\frac{\rho_{e}}{a_{e}}p^{(1)\prime}_{E}+\nonumber\\
\frac{\rho_{P}^p-\rho_{P}}{\Delta t} \Delta x -\rho_{e}u_{e}^{*}+\rho_{w}u_{w}^{*}.\label{eq:corpres}
\end{eqnarray}
    \item Calculate the velocity $u$ using Eq.~(\ref{eq:corvel}). Do not correct the pressure $p^{(1)}$,
    $p^{(1)\prime}$ is used to correct only the velocity field, the pressure being computed
    by Eq.~(\ref{eq:corpres}).
    \item Solve the energy equation for $T$ using the obtained $u$ values.
    \item Calculate the density distribution and $p^{(0)}$ via Eqs.~(\ref{conserv},\ref{EOS3}).
    \item Return to step \ref{st2} and repeat until the converged
solution is obtained.
\end{enumerate}
It has to be noted that whereas the fractional step PISO algorithm \cite{HAMIOUA} is successful in resolving
the equations (\ref{contf}-\ref{EOSf}) on the acoustic time scale, it is not the case when the acoustic filtering
method is used. Due to the different meanings of pressure (see the subsection \ref{Acc}) in the momentum
equation (involving $p^{(1)}$) and in the energy equation (involving $p^{(0)}$), it appears that only an
iterative algorithm can correctly couple the thermodynamic field and the velocity field, the PISO algorithm
leading to unstable solutions.

\end{document}